\documentclass[aps,prb,twocolumn,amssymb,amsmath,superscriptaddress]{revtex4-1}

\usepackage{bm}
\usepackage{color}
\usepackage{graphicx}
\usepackage{dcolumn}
\usepackage{graphicx}
\usepackage[colorlinks=true, letterpaper=true, pdfstartview=FitV, linkcolor=blue, citecolor=blue, urlcolor=blue]{hyperref}
\usepackage[normalem]{ulem}
\usepackage{amsmath}
\usepackage{epstopdf}
\usepackage{multirow}
\usepackage{slashed}
\usepackage{float}
\usepackage{pifont}

\makeatletter
\NAT@superfalse
\setcitestyle{numbers,square,comma,sort&compress}
\renewcommand\@biblabel[1]{#1.}
\makeatother

\begin{document}
	
	\title{Strain induced magnetic phase transition and anomalous transport phenomena in RuO$_2$ and MnF$_2$}
	
	\author{Xiuxian Yang}
	\email{xiuxian$_$yang@outlook.com}
	\affiliation{Jiangsu Key Laboratory of Extreme Multi-Field Materials Physics, School of Physics and Electronic Engineering, Jiangsu Normal University, Xuzhou 221116, China}

	\author{Zhangqi Wu}
	\affiliation{Jiangsu Key Laboratory of Extreme Multi-Field Materials Physics, School of Physics and Electronic Engineering, Jiangsu Normal University, Xuzhou 221116, China}

	\author{Xiangju Wang}
	\affiliation{Centre for Quantum Physics, Key Laboratory of Advanced Optoelectronic Quantum Architecture and Measurement (MOE), School of Physics, Beijing Institute of Technology, Beijing, 100081, China}
	\affiliation{Beijing Key Lab of Nanophotonics \& Ultrafine Optoelectronic Systems, School of Physics, Beijing Institute of Technology, Beijing, 100081, China}
	
	\author{Shifeng Qian}
	\email{qiansf@ahnu.edu.cn}
	\affiliation{Anhui Province Key Laboratory for Control and Applications of Optoelectronic Information Materials, Department of Physics, Anhui Normal University, Anhui, Wuhu 241000, China}	

	\author{Ping Yang}
	\affiliation{Department of Physics, Tianjin Renai College, Tianjin 301636, China}
	
	\author{Xiaodong Zhou}
	\affiliation{School of Physical Science and Technology, Tiangong University, Tianjin 300387, China}


	\author{Jian Hao}
	\email{jian$_$hao@jsnu.edu.cn}
	\affiliation{Jiangsu Key Laboratory of Extreme Multi-Field Materials Physics, School of Physics and Electronic Engineering, Jiangsu Normal University, Xuzhou 221116, China}

	

	\author{Wanxiang Feng}
	\affiliation{Centre for Quantum Physics, Key Laboratory of Advanced Optoelectronic Quantum Architecture and Measurement (MOE), School of Physics, Beijing Institute of Technology, Beijing, 100081, China}
	\affiliation{Beijing Key Lab of Nanophotonics \& Ultrafine Optoelectronic Systems, School of Physics, Beijing Institute of Technology, Beijing, 100081, China}

	\author{Yinwei Li}
	\affiliation{Jiangsu Key Laboratory of Extreme Multi-Field Materials Physics, School of Physics and Electronic Engineering, Jiangsu Normal University, Xuzhou 221116, China}
	
	\date{\today}
	
	\begin{abstract}
	Collinear antiferromagnets with broken time-reversal symmetry have emerged as a fertile platform for spintronics.  Using a general tight-binding model and first-principles calculations, we show that strain engineering provides a simple route to control magnetic phase transition and activate transverse responses in representative altermagnets RuO$_2$ and MnF$_2$.  For pristine RuO$_2$ and MnF$_2$ with N\'eel vector $\mathbf{n}\parallel$ [001], symmetry constrains the off-diagonal elements of the Hall conductivity tensor to vanish, thereby forbidding anomalous transport and magneto-optical responses.  Shear strain applied along the $ac$ direction preserves the spin symmetry relating the two spin-opposite magnetic sublattices and therefore maintains the altermagnetic phase.  By contrast, shear strain applied along the $ab$ direction breaks this spin symmetry and drives a transition from an altermagnetic phase to a partially compensated ferrimagnetic phase in metallic RuO$_2$ and to a fully compensated ferrimagnetic phase in semiconducting MnF$_2$.  In addition, the lowered symmetry enables finite anomalous Hall, anomalous Nernst, and anomalous thermal Hall conductivities, as well as magneto-optical rotation angles, which are prohibited in the pristine systems.  These responses exhibit a clear strain dependence and become progressively stronger as the strain amplitude increases.  Our results establish strain engineering as an effective route to manipulate magnetic phases and functional responses in unconventional antiferromagnets, thereby expanding opportunities for antiferromagnetic spintronics and magneto-optical applications.
	
	\end{abstract}
	
	\maketitle
	
	\section{Introduction}\label{intro}

	Against the backdrop of rapidly developing spintronics, collinear antiferromagnets have attracted particular attention because they offer a highly promising platform for antiferromagnetic spintronic applications.  In these systems, the magnetic moments are strictly antiparallel and collinearly aligned, which have several key advantages over ferromagnets, including insensitivity to external magnetic-field perturbations~\cite{Marti2014,Wadley2016}, ultrafast spin dynamics~\cite{Andrei2010}, and high-frequency spin precession~\cite{Tzschaschel2017,Kampfrath2011,Smejkal_2022_NatRevMater,DaiBQ2025_AFM}.  Owing to these appealing properties, collinear antiferromagnets have become an important platform for exploring novel spin-dependent phenomena and device functionalities.  However, conventional collinear antiferromagnets suffer from spin degeneracy in their electronic band structure, lacking ferromagnet-like responses such as transverse transport phenomena or giant magnetoresistance.  Recently, a new class of collinear antiferromagnets, known as altermagnets (AMs), has been identified both theoretically and experimentally.  AMs combine two seemingly incompatible characteristics: a vanishing net magnetization, as in conventional antiferromagnets, and anisotropic nonrelativistic spin splitting in the electronic band structure, reminiscent of ferromagnets~\cite{Libor2020_SciAdv,ZhouXD_2021_PRB,Libor2022_PRX_a,Libor2022_PRX_b,BaiL_2024_AFM,Reimers_2024_NC,Amin_2024_Nature,Hariki_2024_PRL,Jungwirth_2025_Newton,Song_2025_NatRevMater,Guo_2025_AM}.  Unlike conventional antiferromagnets, in which spin-opposite magnetic sublattices are usually related by inversion or translation symmetry, the defining feature of AMs is that these sublattices are connected by crystallographic rotations or mirror symmetries~\cite{Libor2022_PRX_a,Libor2022_PRX_b,Liu_2022_PRX,ZhouXD_2021_PRB,Chen_2024_PRX,Song_2025_NatRevMater}.  This distinctive symmetry origin gives rise to unconventional alternating spin splitting in momentum space and thereby enables a variety of ferromagnet-like phenomena in AMs, including spin-splitting torque and spin-transfer torque effects~\cite{Rafael_2021_PRL,Bai_2022_PRL,Karube_2022_PRL,Vakili_2025_PRL}, spin-neutral currents~\cite{Shao_2021_NC}, tunnelling magnetoresistance~\cite{Libor_2022_PRXc,Noh_2025_PRL}, Andreev reflection~\cite{Sun_2023_PRB,Papaj_2023_PRB}, topological superconductivity~\cite{Li_2023_PRB,Ghorashi_2024_PRL,Zhu_2023_PRB,Li_2024_PRB}, anomalous Hall effect (AHE)~\cite{Libor2020_SciAdv,Feng_2022_NE,Lee_2024_PRL,ZhouZY_2025_Nature,Takagi_2025_NatMat}, anomalous thermal transport~\cite{Badura_2025_NatCommun,ZhouXD_2024_PRL,YangXX_2025_PRB}, and magneto-optical effects~\cite{ZhouXD_2021_PRB,Hariki_2024_PRL,Olena_2024_SciAdv}.

	Concurrently, another related class of spin-split collinear antiferromagnets has also attracted renewed attention.  
	This concept can be traced back to Pekar and Rashba in 1964~\cite{PekarRashba1965}.  
	When no crystal symmetry relates the spin-opposite magnetic sublattices, their local magnetization distributions may differ in shape while remaining globally compensated~\cite{YuanLD2024_PRL}.  In 1995, van Leuken and de Groot first proposed fully compensated ferrimagnets (fFiMs) in half-metallic Heusler alloys~\cite{Leuken_1995_PRL}, which exhibit zero magnetization and isotropic nonrelativistic spin splitting.  In this situation, provided that at least one spin channel is gapped and the orbital magnetic moment is quenched, the net magnetization of a stoichiometric compound is restricted by the Luttinger theorem to take integer values of Bohr magnetons~\cite{Luttinger1960_PR1,Luttinger1960_PR2,Mazin_2022_PRX}.  When the numbers of spin-up and spin-down electrons are equal, the net magnetization becomes exactly zero and is robust against small perturbations, such as pressure or stress.  However, the number of known fFiMs remains much smaller than that of AMs.  Previous studies of fFiMs have largely focused on three-dimensional materials with sophisticated crystal structures, such as Heusler alloys, transition-metal rare-earth alloys, and organic compounds~\cite{Stinshoff_2017_AIPAdv,Stinshoff_2017_PRB,Midhunlal_2019_JMMM,ZhouHA_2021_JPCJ,Semboshi_2022_SciRep,Seredina_2022_JMMM,Jonathan_2025_SciAdv,Finley_2020_APL}.  Benefiting from recent advances in experimental techniques~\cite{Burch_2018_Nature,Mak_2019_NatRevPhys,Gibertini_2019_NatNano,Gong_2019_Science}, several theoretical works have proposed realizing fFiMs in two-dimensional materials through magnetic heterostructures~\cite{Guo_2025_Arxiv,Gao_2025_CPL,Lishuo_2025JPCL,Zhang_2025_Nanoscale,Guo_2026_PCCP,LiuYC_2025_PRL}.  Nevertheless, despite significant efforts, the realization of fFiMs remains challenging, especially because the synthesis and control of such magnetic heterostructures are experimentally demanding.  Therefore, a simple and controllable experimental route toward fFiMs is urgently needed.

	In this work, we develop a general tight-binding (TB) model for AMs on a rectangular lattice and show that strain engineering can drive a magnetic phase transition from altermagnetism to partially compensated ferrimagnetic phase in metals and fully compensated ferrimagnetic phase in semiconductors.  Guided by this model, and taking RuO$_2$ and MnF$_2$ as representative AMs, we systematically investigate the strain-induced transverse transport responses, including the AHE, anomalous Nernst effect (ANE), and anomalous thermal Hall effect (ATHE), as well as magneto-optical effects, using state-of-the-art first-principles calculations.  We note that the correlation- and volume-sensitive nonmagnetic--magnetic transition in bulk RuO$_2$ has recently been discussed in Ref.~\cite{Hou_RuO2_NM_AM}.  In contrast, the present work focuses on shear-strain-induced symmetry breaking and the resulting transverse response functions in RuO$_2$ and MnF$_2$.  We demonstrate that the magnetic phase transition is strongly anisotropic with respect to the strain direction, occurring only for strain applied along a specific crystallographic axis.  In the pristine unstrained crystal, these anomalous transport responses are prohibited by symmetry, whereas strain engineering breaks the relevant symmetries and thus renders them finite.  In addition, the strain-induced anomalous transport responses exhibit a pronounced dependence on strain strength, and increasing strain provides an effective means to tune their magnitudes.

	\section{Theory and computational details}\label{method}
	
	The anomalous Hall effect (AHE) has a history of more than a century~\cite{Hall1881} and remains a longstanding central topic in condensed matter physics, playing an indispensable role in elucidating the microscopic origin of magnetism~\cite{Nagaosa2010,Smejkal_2022_NatRevMater}.  The AHE refers to a transverse charge current induced by a longitudinal electric field in the absence of an external magnetic field.  Within the Berry-phase theory and the Kubo formalism, the intrinsic anomalous Hall conductivity (AHC)~\cite{Yao_2004_PRL} can be written as 
	\begin{eqnarray}\label{eq:int_AHC}
		\sigma_{ij} = -\frac{e^2}{\hbar}\sum_n\int \frac{d^3k}{(2\pi)^3}\Omega^n_{ij}(\textbf{k})f_{n\textbf{k}}, 
	\end{eqnarray}
	where $f_{n\textbf{k}}$ is the Fermi-Dirac distribution function, and $\Omega^n_{ij}(\textbf{k})$ is the band-resolved Berry curvature, defined as
	\begin{eqnarray}\label{eq:Berry}
		\Omega^n_{ij} = -\sum_{n'\neq n}\frac{2\rm Im\left[\left\langle \psi_{n\textbf{k}}| \hat{v}_i| \psi_{n'\textbf{k}} \right\rangle \left\langle \psi_{n'\textbf{k}}| \hat{v}_j| \psi_{n\textbf{k}} \right\rangle\right] }{(\omega_{n'\textbf{k}}-\omega_{n\textbf{k}})^2}.
	\end{eqnarray}
	Here, $\{i, j\} \in \{x, y, z\}$ denote the Cartesian coordinates, $\hat{v}_{i, j}$ are the velocity operators, $\psi_{n\textbf{k}}$ and $\varepsilon_{n\textbf{k}} = \hbar\omega_{n\textbf{k}}$ are the eigenvector and eigenvalue at the band index $n$ and momentum $\textbf{k}$, respectively.

	A transverse charge current and transverse thermal current can also be driven by a longitudinal temperature gradient, known as the anomalous Nernst effect (ANE) and anomalous thermal Hall effect (ATHE), respectively.  The ANE and ATHE are commonly regarded as the thermoelectric counterpart and thermal analog of the AHE, respectively.  These three effects can be interconnected through the anomalous transport coefficients in the generalized Landauer-B\"uttiker formalism~\cite{ashcroft1976solid,Houten1992,behnia2015fundamentals},
	\begin{equation}\label{eq:LB}
		R^{(n)}_{ij}=\int^\infty_{-\infty}(E-\mu)^n\left(-\frac{\partial f}{\partial E}\right)\sigma_{ij}(E)dE,
	\end{equation}
	where $\mu$ is the chemical potential, $f = 1/[\textnormal{exp}((E-\mu)/k_{B}T) + 1]$ represents the Fermi-Dirac distribution function, and $\sigma_{ij}$ is the AHC at zero temperature. Then the temperature-dependent AHC, anomalous Nernst conductivity (ANC), and anomalous thermal Hall conductivity (ATHC) read
	\begin{eqnarray}
	    \sigma_{ij}^T & = &R^{(0)}_{ij}, \label{eq:TAHC} \\
		\alpha_{ij}^T & = &-R^{(1)}_{ij}/eT, \label{eq:ANC} \\
		\kappa_{ij}^T & = & R^{(2)}_{ij}/e^2T. \label{eq:ATHC}
	\end{eqnarray}

	Magneto-optical effects constitute another important class of phenomena in condensed matter physics, which provide a powerful probe of the microscopic interaction between light waves and magnetic media.  Among magneto-optical effects, two representative effects are the magneto-optical Kerr effect (MOKE)~\cite{Kerr_1877} and the magneto-optical Faraday effect (MOFE)~\cite{Faraday_1846}, which describe the rotations of the polarization plane of the linearly polarized light reflected from and transmitted through magnetic media, respectively.  We quantify the magneto-optical performance of a given magnetic medium by the rotation of the polarization plane with respect to the incident light, namely the Kerr and Faraday rotations ($\theta_K$ and $\theta_F$).  Additionally, the rotation angle and the ellipticity ($\varepsilon_K$ and $\varepsilon_F$) are often combined into the so-called complex Kerr and Faraday angles ($\varTheta_K$ and $\varTheta_F$), which can be written as~\cite{Guo_1995_PRB,Ravindran_1999_PRB,Feng_2015_PRB,Wimmer_2019_PRB,ZhouXD_2019_PRB}
	\begin{equation}\label{eq:MOK}
		\varTheta_K = \theta_K + i\varepsilon_K = -\frac{\nu_{ijk}\sigma_{ij}(\omega)}{\sigma_0(\omega)\sqrt{1+i(4\pi/\omega)\sigma_0(\omega)}},
	\end{equation}
	and
	\begin{equation}\label{eq:MOF}
		\begin{aligned}
		\varTheta_F &= \theta_F + i\varepsilon_F = \nu_{ijk}(n_+ - n_-)\frac{\omega d}{2c} \\
		&\approx  -\frac{\nu_{ijk}\sigma_{ij}}{\sqrt{1+i(4\pi/\omega)\sigma_0}}\frac{2\pi d}{c},
		\end{aligned}
	\end{equation}
	\begin{equation}\label{eq:Refractive}
		n_\pm = \sqrt{1+4\pi i/\omega(\sigma_0 \pm i\nu_{ijk}\sigma_{ij})}.
	\end{equation}
	Here, the $\nu$, $c$, $d$, and $\omega$ are the Levi-Civita symbol, speed of light in the vacuum, thickness of a thin film, and the frequency of incident light, respectively.  $\sigma_{ij}$ denotes the off-diagonal element of the optical conductivity tensor for a magnetic medium, and $\sigma_0 = (\sigma_{ii} + \sigma_{jj})/2$ $(i\neq j)$.  It should be noted that for bulk materials, the thickness $d$ need not be specified, since the Faraday angles are given in unit of deg/$\mu$m in our calculations.

	\begin{figure*}[t]
		\includegraphics[width=2\columnwidth]{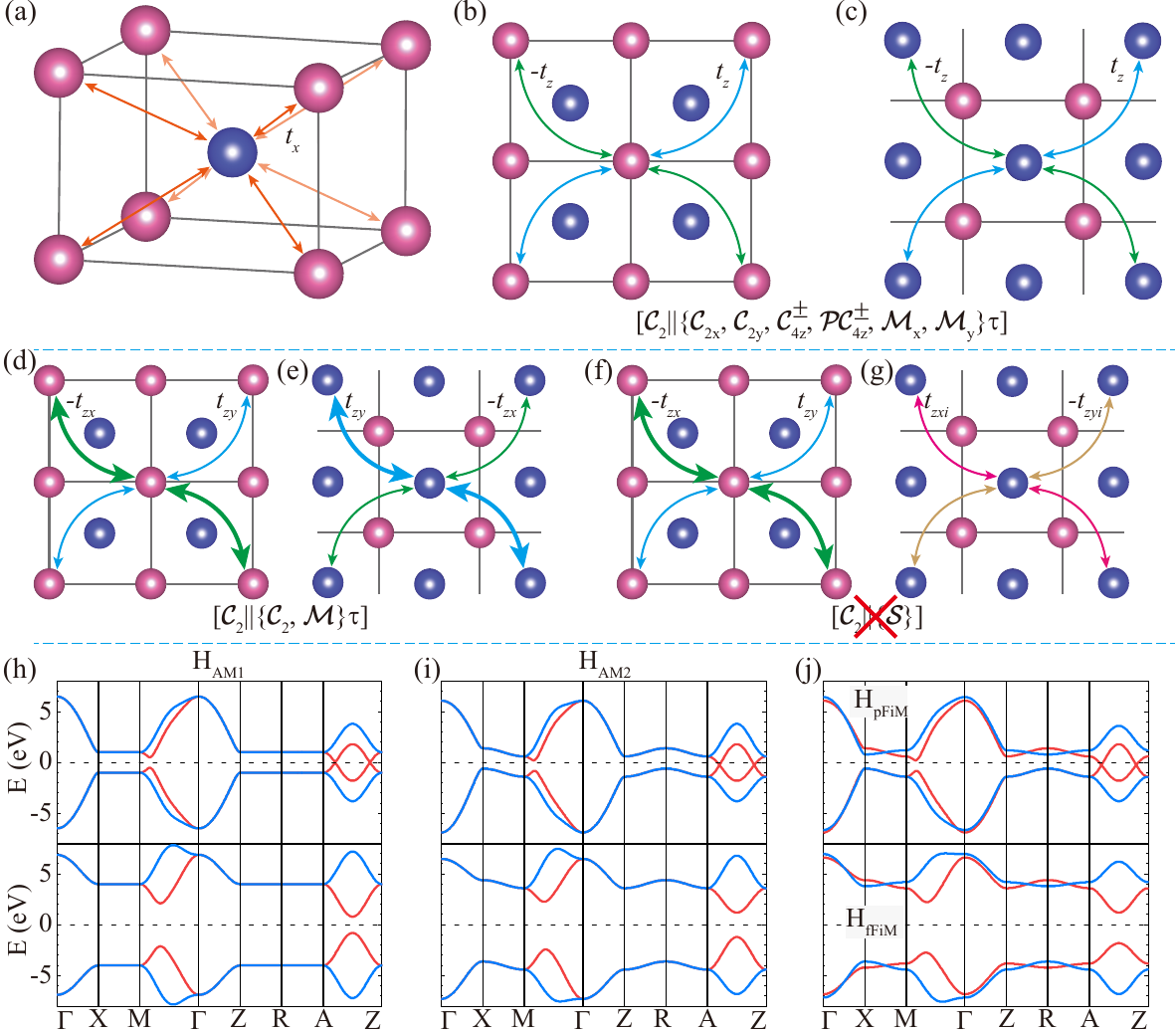}
		\caption{(Color online)  Crystal structures, relevant hopping processes, and spin-resolved band structures of the TB models $H_{AM1}$, $H_{AM2}$, and $H_{FiM}$ (or $H_{fFiM}$), where the red and blue colors denote the two magnetic sublattices.  (a) Three dimensional lattice illustration of the $t_x$.  (b-c) Top view of the lattice showing the $t_z$.  (d) and (e) Top view of the lattice showing the $t_{zx}$ and $-t_{zy}$.  (f) and (g) Top view of the lattice showing the $t_{zx}$, $-t_{zy}$, $t_{zxi}$, and $-t_{zyi}$.  In each of (h)-(j), the upper panel corresponds to the metallic case, while the lower panel corresponds to the semiconducting case. Accordingly, after the strain-induced symmetry breaking, the metallic system evolves into a partially compensated ferrimagnetic phase, whereas the semiconducting system becomes a fully compensated ferrimagnetic phase.}
		\label{fig:model}
	\end{figure*}

	As follows from Eqs.~\ref{eq:MOK}--\ref{eq:Refractive}, it is evident that the magneto-optical effects are closely related to the optical conductivity tensor, which can be calculated using the Kubo-Greenwood formula on the basis of Wannier functions~\cite{Mostofi_2008_CPC,Yates_2007_PRB},
	\begin{equation}\label{eq:OPC}
		\begin{aligned}
		\sigma_{ij}(\omega) &= \sigma'_{ij}(\omega) + i\sigma''_{ij}(\omega) = \frac{i e^2 \hbar}{N_k \Omega_c}\sum_{\mathbf{k}} \sum_{n,m}\frac{f_{m\mathbf{k}} - f_{n\mathbf{k}}}{\varepsilon_{m\mathbf{k}} - \varepsilon_{n\mathbf{k}}} \\
		&\times \frac{\langle \psi_{n\mathbf{k}} | \hat{v}_i | \psi_{m\mathbf{k}} \rangle \langle \psi_{m\mathbf{k}} | \hat{v}_j | \psi_{n\mathbf{k}} \rangle }{\varepsilon_{m\mathbf{k}} - \varepsilon_{n\mathbf{k}} - (\hbar\omega + i\eta)
		},
		\end{aligned}
	\end{equation}
	where the subscripts $'$ and $''$ denote the real and imaginary parts of $\sigma_{ij}$, $\Omega_c$ is the cell volume, $N_k$ is the total number of $k$-points used for sampling the first Brillouin zone, and $\eta$ is an adjustable energy smearing parameter.  In our work, a $301\times 301\times 401$ $k$-mesh and an energy smearing parameter $\eta = 0.1 eV$ were used for RuO$_2$.

	The first-principles calculations were performed using the Vienna \textit{ab initio} simulation package (VASP)~\cite{Kresse1993,Kresse1996}.  The exchange-correlation functional was treated within the generalized gradient approximation (GGA) using the Perdew-Burke-Ernzerhof parameterization~\cite{Perdew1996}.  The plane-wave cutoff energy was set to 500 eV.  The structures were fully relaxed until the force and energy convergence criteria were smaller than 0.001 eV/\AA{} and 10$^{-7}$ eV, respectively.  For the self-consistent calculations, a Monkhorst-Pack k-point mesh of 16$\times$16$\times$24 was used.  The GGA+U method~\cite{Anisimov_LDAU1991,Dudarev_LDAU1998} was employed to account for the Coulomb correction of the $d$ orbitals of Ru atoms with U = 2.0 eV and J = 0.4 eV, and of Mn atoms with U = 4.0 eV and J = 0 eV~\cite{YangXX_2025_PRB}.  To construct maximally localized Wannier functions, $s$, $p$, $d$ orbitals of Ru and $p$ orbitals of O atoms were projected onto a uniform $k$-mesh of 8$\times$8$\times$8 using the \textsc{wannier90} package~\cite{Arash2008}.  The AHC was calculated with an energy interval of 0.1 meV using the WannierBerri code~\cite{wannierberri} on an ultra-dense $k$-mesh of 301$\times$301$\times$401.  The phonon spectrum of strained RuO$_2$ was calculated using the finite-difference approach implemented in the PHONOPY package~\cite{phonopy_phono3py_JPCM,phonopy_phono3py_JPSJ}, with a 3$\times$3$\times$3 supercell.

	\section{Results and discussion}\label{results}

	Altermagnetic order is essentially a form of magnetic crystal order jointly determined by the N\'eel vector ($\mathbf{n}$) and the underlying crystal symmetry.  Previous studies have mainly focused on manipulating altermagnetic symmetry by reorienting the $\mathbf{n}$~\cite{Feng_2022_NE,Lee_2024_PRL,WangM_2023_NC,HanL_2024_SciAdv}, whereas recent works have begun to explore the role of strain in tuning altermagnetic states~\cite{chen_AM2025,strainLeiBC2025,Jeong2025,Johnathas2025,ZhouZY_2025_Nature}.  To elucidate the effect of strain on the magnetic and electronic structures of AMs, we perform an analysis combining symmetry considerations with a minimal tight-binding (TB) model, taking metallic RuO$_2$ and semiconducting MnF$_2$ as representative examples.  In pristine structures, the magnetic atoms (Ru or Mn) occupy sites with $D_{2h}$ symmetry, while the overall crystallographic point group is $D_{4h}$.  The spin-opposite sublattices are connected and protected by symmetry operations such as the four-fold rotational symmetry ($C_{4z}$) and the two-fold rotational symmetry ($C_{2y}$).  A minimal TB model capturing the essential physics of AMs requires only two sublattices~\cite{Roig2024_PRB}.  Let the sublattice degrees of freedom for the magnetic ions in the unit cell be described by the Pauli matrices $\tau_i$, while the spin degrees of freedom are represented by the Pauli matrices $\sigma_i$.  We also assume a single relevant orbital degree of freedom at each lattice site.
	
	The crystal structure and the relevant hopping pathways for the minimal model are schematically depicted in Figs.~\ref{fig:model}(a-c). The general Hamiltonian for this minimal model exhibiting altermagnetism takes the form:
	\begin{equation}
	H_{AM1}=\lambda \sigma_z \tau_z +t_{x, \mathbf{k}} \tau_x+t_{z, \mathbf{k}} \tau_z,
	\end{equation}
	where $t_{x, \mathbf{k}}=8t_x \cos \frac{k_x}{2} \cos \frac{k_y}{2} \cos \frac{k_z}{2}$ and $t_{z, \mathbf{k}}=4t_{z} \sin k_x \sin k_y$, and $\lambda$ represents the strength of the exchange field.  By adopting the hopping parameters $\lambda=t,t_x = 0.8t$ and $t_z = 0.7t$, the resulting band structure, as shown in Fig.~\ref{fig:model}(h), clearly exhibits the characteristic spin-splitting features of a $d$-wave altermagnet.
	
	\begin{figure*}[t]
		\includegraphics[width=2\columnwidth]{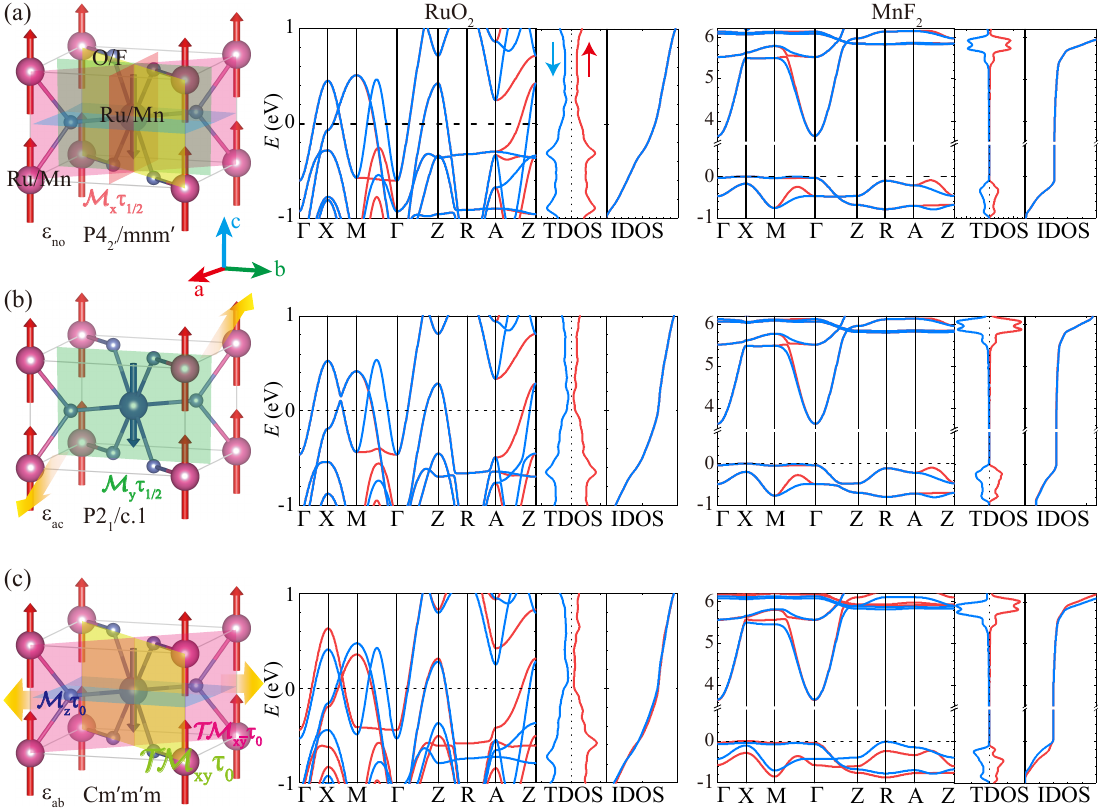}
		\caption{(Color online)  Schematic crystal structures, together with the corresponding nonrelativistic band structures, total density of states (TDOS), and integrated density of states (IDOS), of RuO$_2$ and MnF$_2$ with the N\'eel vector $\mathbf{n}\parallel$ [001].  Panels (a), (b), and (c) correspond to the unstrained state ($\varepsilon_{no}$), shear strain applied along the $ac$ direction ($\varepsilon_{ac}$), and along the $ab$ direction ($\varepsilon_{ab}$).  In each panel, the left parts shows the schematic crystal structure, while the middle and right part show the nonrelativistic band structure, TDOS, and IDOS of RuO$_2$ and MnF$_2$, respectively.  In (b) and (c), the band structures and DOS are calculated for a strain strength of $4\%$.  The red and blue lines denote the spin-up and spin-down channels, respectively.  For the $\varepsilon_{no}$ (a), the magnetic space group is $P4_{2'}/mnm'$.  The colored planes indicate the symmetry operations relevant to the anomalous transport and magneto-optical responses, including the mirror operation $\mathcal{M}_z\tau_0$, the glide-mirror operations $\mathcal{M}_x\tau_{\frac{1}{2}}$ and $\mathcal{M}_y\tau_{\frac{1}{2}}$, and the combined operations $\mathcal{TM}_{xy}\tau_0$ and $\mathcal{TM}_{x\bar{y}}\tau_0$.  For the case of $\varepsilon_{ac}$ (b), the corresponding magnetic space group is $P2_1'/c.1$, and the remaining symmetry operation is $\mathcal{M}_y\tau_{\frac{1}{2}}$.  For the case of $\varepsilon_{ab}$ (c), the corresponding magnetic space group is $Cm'm'm$, and the remaining symmetry operations are $\mathcal{M}_z\tau_0$, $\mathcal{TM}_{xy}\tau_0$, and $\mathcal{TM}_{x\bar{y}}\tau_0$.  The phonopy spectrum of the strained states are shown in the Fig.~\textcolor{blue}{S1}~\cite{SuppMater}, confirming the dynamical stability of the strained structures.}
		\label{fig:structure}
	\end{figure*}

	When strain is applied, several distinct scenarios can arise depending on the strain direction.  First, if the applied strain does not break the core symmetries connecting the two sublattices (e.g., uniaxial strain along the $c$-axis), the magnetic phase remains unchanged.  Second, if the strain breaks the $C_{4z}$ symmetry but preserves either the $C_{2x}$ or $C_{2y}$ symmetry, the two spin-opposite sublattices remain related by these surviving two-fold rotations. Consequently, despite the structural symmetry reduction, the altermagnetic state remains protected and the magnetic phase is preserved.  In this scenario, the relevant TB hoppings are illustrated in Figs.~\ref{fig:model}(a), \ref{fig:model}(d), and \ref{fig:model}(e). The modified TB Hamiltonian takes the form:
	\begin{equation}
	H_{AM2}=\lambda \sigma_z \tau_z +t_{x, \mathbf{k}} \tau_x+t_{za, \mathbf{k}} \tau_z+t_{zb, \mathbf{k}},
	\end{equation}
	where $t_{za, \mathbf{k}}=(t_{zx}-t_{zy}) \sin k_x \sin k_y$ and $t_{zb, \mathbf{k}}=(t_{zx}+t_{zy}) \sin k_x \sin k_y$.
	Using the parameters $t_{zx} = 0.8t$ and $t_{zy} = 0.6t$, the calculated band structure still displays the hallmark alternating spin splitting of $d$-wave altermagnetism.  Finally, if the strain further breaks the remaining $C_{2x}$ or $C_{2y}$ symmetries, a fundamental magnetic phase transition occurs.  Without any symmetry operations left to map the two opposite-spin sublattices onto one another, the degeneracy is fully lifted, and the system transitions into a partially compensated ferrimagnetic phase in metals and fully compensated ferrimagnetic phase in semiconductors. The relevant hoppings for this symmetry-broken TB model are shown in Figs.~\ref{fig:model}(a), \ref{fig:model}(f), and \ref{fig:model}(g).  The other parameters are kept consistent with $H_{AM2}$.  Utilizing the parameters ($t_{zxi} = 0.8t$ and $t_{zyi} = 0.6t$), the band structure manifestly exhibits the isotropic nonrelativistic spin splitting, consistent with partially compensated ferrimagnetism in metals and fully compensated ferrimagnetism in semiconductors.

	Having established the strain-driven phase evolution within the minimal TB model, we now turn to realistic materials and examine whether these symmetry-governed strain effects can be realized in first-principles calculations.  The pristine bulk RuO$_2$ and MnF$_2$ crystallize in the rutile-type tetragonal structure with space group $P4_2/mnm$, in which the magnetic atoms are located at the centers of slightly elongated RuO$_6$ and MnF$_6$ octahedra, respectively, as shown in Fig.~\ref{fig:structure}(a).  The optimized lattice constants are $a = b = 4.54$ \AA{} and $c = 3.11$ \AA{} for RuO$_2$, and $a = b = 4.95$ \AA{} and $c = 3.34$ \AA{} for MnF$_2$, in good agreement with previous studies~\cite{Baur_1971_ActaCrystB,Berlijn_2017_PRL,YanLD_2020_PRB}.  The N\'eel vector is defined as $\mathbf{n} = (\mathbf{n}_{1} - \mathbf{n}_{2})/2$, where $\mathbf{n}_{1}$ and $\mathbf{n}_{2}$ denote the magnetic moments on the two magnetic sublattices.  Experiments have established that the N\'eel vectors in both RuO$_2$ and MnF$_2$ are aligned along the [001] direction~\cite{Baur_1971_ActaCrystB,Berlijn_2017_PRL,ZhuZH_2019_PRL,YanLD_2020_PRB,Feng_2022_NE}.

	The pristine RuO$_2$ and MnF$_2$ are both AMs, in which two spin-opposite magnetic sublattices are related by the spin-space group symmetry $[\mathcal{C}_2\parallel \{\mathcal{O}\}]$, where $\mathcal{C}_2$ denotes spin rotation and ${\mathcal{O}}$ represents a rotational or mirror symmetry.  The corresponding spin-space group contains the following symmetry operations,
	\begin{eqnarray}
	\varepsilon_{no}: [\mathcal{E} || \{\mathcal{E}, \mathcal{P}, \mathcal{C}_{2z}, \mathcal{C}_{2xy}, \mathcal{C}_{2x\bar{y}}, \mathcal{M}_{z}, \mathcal{M}_{xy}, \mathcal{M}_{x\bar{y}}\}] + \notag \\
	\ [\mathcal{C}_{2}||\{\mathcal{C}_{2x}, \mathcal{C}_{2y}, \mathcal{C}_{4z}^+, \mathcal{C}_{4z}^-, \mathcal{PC}_{4z}^+, \mathcal{PC}_{4z}^-, \mathcal{M}_{x}, \mathcal{M}_{y}\}\tau]. \label{eq:nostrain}
	\end{eqnarray}
	Here, $\mathcal{E}$, $\mathcal{P}$, $\mathcal{M}$, and $\tau$ represent identity, space inversion, mirror, and translation operations, respectively.  These symmetry operations imply that the spin splitting of bands alternates across two perpendicular wave vectors in the $k_z$ plane, leading to a planar $d$-wave spin-momentum-locking band structure.  This picture is directly confirmed by the first-principles calculations shown in Fig.~\ref{fig:structure}(a).  The red and blue lines denote the spin-up and spin-down channels, respectively.  One can see that the bulk RuO$_2$ is metallic, with both valence and conduction bands crossing the Fermi level, whereas bulk MnF$_2$ is semiconducting with a large band gap of 3.63 eV.  A pronounced momentum-dependent nonrelativistic spin splitting is clearly resolved along the $\rm{M}-\Gamma$ and $\rm{A-Z}$ high-symmetry paths, constituting a characteristic fingerprint of altermagnetism.  In particular, the sizable spin splitting along the $\rm{M}-\Gamma$ path provides a robust experimental fingerprint for identifying and confirming the long-range magnetic order~\cite{LinZH2024}.  Moreover, the total density of states (TDOS) and integrated density of states (IDOS) show that the spin-up and spin-down channels are exactly compensated, yielding a strictly zero net magnetization.

	We now examine whether the strain-driven phase evolution predicted by the minimal TB model can be realized in realistic material.  We first consider the shear strain applied along the $ac$ direction $\varepsilon_{ac}$, seeing Fig.~\ref{fig:structure}(b), for which the nonrelativistic spin-space group incorporates the following symmetry operations,
	\begin{eqnarray}
		\varepsilon_{ac}: &[\mathcal{E}||\{\mathcal{E}, \mathcal{P}\}] + [\mathcal{C}_2||\{{\mathcal{C}_{2y}, \mathcal{M}_y}\}\tau].\label{eq:ac_strain} 
	\end{eqnarray}
	The $\mathcal{C}_2$ spin symmetry operation is preserved when $\varepsilon_{ac}$ is applied, which, together with the collinear antiferromagnetic order, suggests that the magnetic phase remains altermagnetic.  This expectation is directly confirmed by the nonrelativistic band structure in Fig.~\ref{fig:structure}(b), which exhibits the characteristic alternating spin splitting together with a compensated TDOS and identical IDOS, giving rise to zero net magnetization. The nonrelativistic band structures for different strain amplitudes applied along the $ac$ direction are provided in the Fig.~\textcolor{blue}{S2} of Supplemental Material (SM)~\cite{SuppMater}.

	For the shear strain applied along the $ab$ direction $\varepsilon_{ab}$, seeing Fig.~\ref{fig:structure}(c), the nonrelativistic spin-space group contains the following symmetry operations,
	\begin{eqnarray}
		\varepsilon_{ab}: &[\mathcal{E}||\{\mathcal{E}, \mathcal{P}, \mathcal{C}_{2z}, \mathcal{C}_{2xy}, \mathcal{C}_{2x\bar{y}}, \mathcal{M}_z, \mathcal{M}_{xy}, \mathcal{M}_{x\bar{y}}\}]. \label{eq:ab_strain}
	\end{eqnarray}
	In contrast to the $\varepsilon_{ac}$ case, the $\mathcal{C}_2$ spin symmetry is completely absent here, indicating that the system is driven out of the altermagnetic phase. In this nonrelativistic spin group, the two spin-opposite magnetic sublattices can no longer be related by any symmetry operation.  As a result, the calculated nonrelativistic band structures in Fig.~\ref{fig:structure}(c) exhibit isotropic spin splitting throughout the first Brillouin zone, signaling the emergence of a ferrimagnetic rather than altermagnetic state.  Importantly, the character of this strain-induced ferrimagnetic phase depends on the electronic nature of the material.  For metallic RuO$_2$, the TDOS of spin-up and spin-down channels are no longer fully compensated, while the IDOS of two spin channels remain very close to each other, giving rise to an almost vanishing net magnetic moment of about $0.002~\mu_{\mathrm{B}}$ per Ru atom.  RuO$_2$ is thus identified as a partially compensated ferrimagnetic metal under $\varepsilon_{ab}$.  By contrast, for semiconducting MnF$_2$, the spin-up and spin-down TDOS are no longer fully compensated due to the isotropic spin splitting.  However, the IDOS of spin-up and spin-down remain integer and exactly equal at the Fermi energy, resulting in a strictly zero net magnetization.  This identifies MnF$_2$ as a fully compensated ferrimagnetic semiconductor under $\varepsilon_{ab}$.  The nonrelativistic band structures for different strain amplitudes applied along the $ab$ direction are provided in the Fig.~\textcolor{blue}{S3}~\cite{SuppMater}.

	\begin{figure*}[t]
		\includegraphics[width=2\columnwidth]{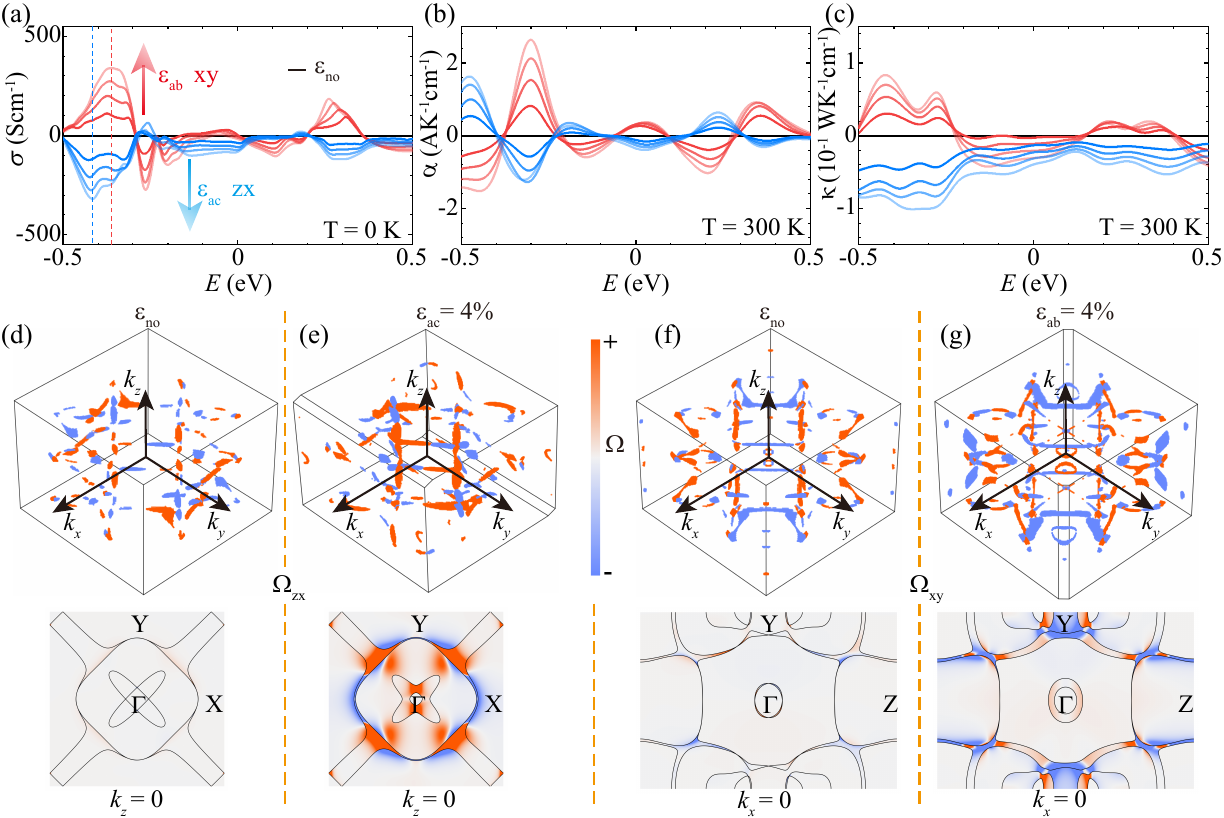}
		\caption{(Color online)  Anomalous transport properties of RuO$_2$ with the N\'eel vector $\mathbf{n}\parallel$ [001].  (a)-(c) Anomalous Hall conductivity ($\sigma$, T = 0 K), anomalous Nernst conductivity ($\alpha$, T = 300 K), and anomalous thermal Hall conductivity ($\kappa$, T = 300 K) as functions of Fermi energy.  The blue and red solid lines indicate strain applied along the $ac$ ($\varepsilon_{ac}$) and $ab$ ($\varepsilon_{ab}$) directions, respectively, and the color gradient from dark to light denotes increasing strain amplitude.  For comparison, the anomalous transport coefficients of the unstrained case ($\varepsilon_{no}$) are also plotted as black solid lines in (a)-(c).  (d-g) Distribution of Berry curvature in the first Brillouin zone.  In each panel, the upper plot shows the distribution of Berry curvature in the three-dimensional Brillouin zone, while the lower plot shows the Berry curvature together with the Fermi surface on the corresponding two-dimensional momentum plane.  To better illustrate the strain-induced changes in anomalous transport, the Berry-curvature distributions for the pristine case ($\varepsilon_{no}$) are also included.  Specifically, (d) and (e) show $\Omega_{zx}$ for $\varepsilon_{no}$ and $\varepsilon_{ac}$, respectively, on the $k_x$ plane, while (f) and (g) show $\Omega_{xy}$ for $\varepsilon_{no}$ and $\varepsilon_{ab}$, respectively, on the $k_z$ plane.}
		\label{fig:AHC}
	\end{figure*}

	Next, we analyze the anomalous transport phenomena in RuO$_2$ and MnF$_2$.  For metallic RuO$_2$, we mainly focus on the AHE, ANE, and ATHE, whereas for semiconducting MnF$_2$, we primarily investigate the MOKE and MOFE.  Before discussing these responses in detail, we first examine the relevant symmetry constraints.  As follows from Eqs.~\ref{eq:int_AHC}--\ref{eq:ATHC}, the finite-temperature AHC, ANC, and ATHC can all be derived from the zero-temperature AHC.  Moreover, the real part of the off-diagonal optical conductivity reduces to the zero-temperature AHC in the dc limit.  These relations indicate that all of the above effects share the same symmetry requirements.  Therefore, it is sufficient to analyze the symmetry requirements of the optical conductivity tensor $\bm{\sigma}$.  Magnetic group theory provides a powerful tool for determining the shape of $\bm{\sigma}$.  In addition, the off-diagonal components of $\bm{\sigma}$ can be regarded as a pseudovector, analogous to spin.  For convenience, we therefore adopt the vector-form notation $\bm{\sigma} = [\sigma^x, \sigma^y, \sigma^z] = [\sigma_{yz}, \sigma_{zx}, \sigma_{xy}]$.

	When $\mathbf{n}$ is aligned along the [001] direction, the pristine ($\varepsilon_{no}$) RuO$_2$ and MnF$_2$ belong to the magnetic space group $P4_{2'}/mnm'$.  This group contains one mirror operation ($\mathcal{M}_z\tau_0$), two glide-mirror operations ($\mathcal{M}_x\tau_{\frac{1}{2}}$ and $\mathcal{M}_y\tau_{\frac{1}{2}}$), and two combined symmetries ($\mathcal{TM}_{xy}\tau_0$ and $\mathcal{TM}_{x\bar{y}}\tau_0$).  The corresponding mirror planes are indicated by different colors in Fig.~\ref{fig:structure}(a).  Here, the $\mathcal{T}$ denotes time-reversal operation.  Since the $\tau$ acts trivially on the $\bm{\sigma}$, i.e., $\tau \bm{\sigma} = \bm{\sigma}$, only the $\mathcal{M}$ and $\mathcal{T}$ need to be considered when determining the symmetry constraints on $\bm{\sigma}$.  Specifically, the $\mathcal{M}_z\tau_0$ operation flips the signs of $\sigma^x$ and $\sigma^y$ while leaving $\sigma^z$ invariant.  The glide mirror $\mathcal{M}_x\tau_{\frac{1}{2}}$ reverses the signs of $\sigma^y$ and $\sigma^z$ but preserves $\sigma^x$.  Similarly, $\mathcal{M}_y\tau_{\frac{1}{2}}$ leaves $\sigma^y$ unchanged while flipping the signs of $\sigma^x$ and $\sigma^z$.   Taken together, these operations force all components of the $\bm{\sigma}$ to vanish, yielding $\bm{\sigma} = [0, 0, 0]$.  Consequently, these effects are symmetry forbidden in pristine RuO$_2$ and MnF$_2$.  Previous studies therefore mainly relied on rotating the magnetic moments to break the relevant symmetries~\cite{Libor2020_SciAdv,Feng_2022_NE,ZhouXD_2024_PRL,ZhouXD_2021_PRB}, which typically requires an external magnetic field.

	When shear strain is applied along the crystallographic $ac$ direction ($\varepsilon_{ac}$) [Fig.~\ref{fig:structure}(b)], all symmetries are broken except $\mathcal{M}_y\tau_{\frac{1}{2}}$, which allows only $\sigma^y \neq 0$ and thus leads to $\bm{\sigma} = [0, \sigma^y, 0]$.  When shear strain is applied along the crystallographic $ab$ direction ($\varepsilon_{ab}$) [Fig.~\ref{fig:structure}(c)], the two glide-mirror operations are broken, leaving only the mirror operation $\mathcal{M}_z\tau_0$ and two combined operations $\mathcal{TM}_{xy}\tau_{\frac{1}{2}}$ and $\mathcal{TM}{x\bar{y}}\tau_{\frac{1}{2}}$.  Both $\mathcal{TM}_{xy}\tau_{\frac{1}{2}}$ and $\mathcal{TM}{x\bar{y}}\tau_{\frac{1}{2}}$ reverse the signs of $\sigma^x$ and $\sigma^y$ while preserving $\sigma^z$. Meanwhile, $\mathcal{M}_z\tau_0$ also flips the signs of $\sigma^x$ and $\sigma^y$ but leaves $\sigma^z$ invariant. As a result, only the $\sigma^z$ component is allowed to be nonzero, namely $\bm{\sigma} = [0, 0, \sigma^z]$.  In summary, $\varepsilon_{ac}$ and $\varepsilon_{ab}$ selectively activate $\sigma^y$ and $\sigma^z$, respectively, thereby making the corresponding AHE, ANE, ATHE, MOKE, and MOFE symmetry allowed.
	


	As implied by the symmetry analysis above, pristine RuO$_2$ forbids all off-diagonal responses, whereas $\varepsilon_{ac}$ and $\varepsilon_{ab}$ selectively activate the $zx$ and $xy$ components, respectively.  The corresponding relativistic band structures including SOC are presented in Figs.~\textcolor{blue}{S4--S5}~\cite{SuppMater}, and the strain-induced anomalous transport responses are shown in Fig.~\ref{fig:AHC} for RuO$_2$.  Figs.~\ref{fig:AHC}(a-c) present the AHC at T = 0 K, as well as the ANC and ATHC at T = 300 K, as a functions of the Fermi energy for $\varepsilon_{no}$ (black solid lines), $\varepsilon_{ac}$ (blue solid lines), and $\varepsilon_{ab}$ (red solid lines).  The AHC at 300 K is plotted in Fig.~\textcolor{blue}{S6}.  As expected, the black curves vanish identically due to the symmetry constraints in pristine RuO$_2$.  Accordingly, finite off-diagonal transport coefficients emerge in both strain configurations, and their overall magnitudes generally increase with increasing strain amplitude.  In the following, we take a strain strength of $4\%$ as a representative example.  Near the E$_F$, the values are ($\sigma_{zx}$, $\alpha_{zx}$, $\kappa_{zx}$) = (-81.5 Sm$^{-1}$, 0.3 AK$^{-1}$m$^{-1}$, -0.04 WK$^{-1}$m$^{-1}$) for $\varepsilon_{ac}$, and ($\sigma_{xy}$, $\alpha_{xy}$, $\kappa_{xy}$) = (-17.0 Sm$^{-1}$, 0.3 AK$^{-1}$m$^{-1}$, -0.03 WK$^{-1}$m$^{-1}$) for $\varepsilon_{ab}$.  Carrier doping can further enhance these signals.  For example, in the hole-doped regime, we obtain ($\sigma_{zx}$, $\alpha_{zx}$, $\kappa_{zx}$) = (-318.7 Sm$^{-1}$, 1.62 AK$^{-1}$m$^{-1}$, 0.10 WK$^{-1}$m$^{-1}$) for $\varepsilon_{ac}$ and ($\sigma_{xy}$, $\alpha_{xy}$, $\kappa_{xy}$) = (340.5 Sm$^{-1}$, 2.63 AK$^{-1}$m$^{-1}$, 0.08 WK$^{-1}$m$^{-1}$) for $\varepsilon_{ab}$.

	To elucidate the microscopic origin of AHE, Figs.~\ref{fig:AHC}(d-g) show the distribution of Berry curvature ($\Omega_{xy}$ and $\Omega_{zx}$) in the three-dimensional Brillouin zone and the corresponding two-dimensional momentum planes.  In the absence of strain, the positive and negative Berry curvature contributions tend to cancel upon integration, leading to vanishing anomalous transport coefficients.  Once shear strain is applied, the distribution of positive and negative Berry curvature becomes unbalanced.  For the $\varepsilon_{ab}$, the negative spots of $\Omega_{xy}$ are obviously larger than the positive ones [Fig.~\ref{fig:AHC}(e)].  According to Eq.~\ref{eq:int_AHC}--\ref{eq:Berry}, where the AHC is proportional to the negative integral of the Berry curvature, this imbalance gives rise to a positive AHC.  By contrast, for the $\varepsilon_{ac}$, the positive spots of $\Omega_{zx}$ dominate over the negative ones [Fig.~\ref{fig:AHC}(g)], resulting in the negative values of AHC.
	

	\begin{figure}
		\includegraphics[width=1\columnwidth]{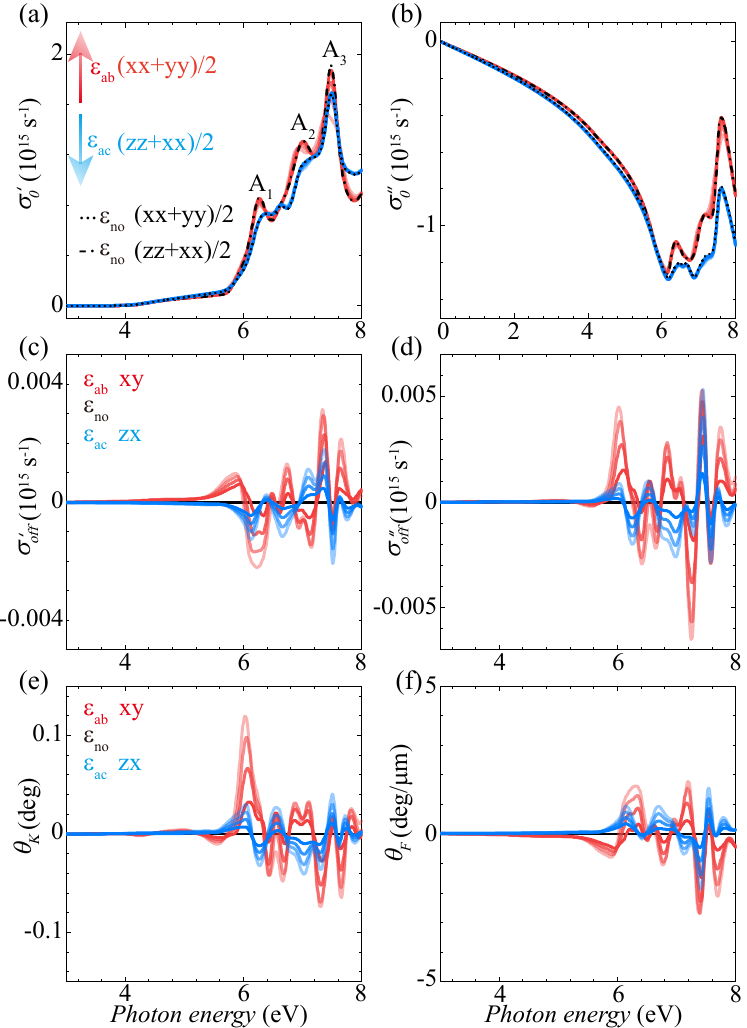}
		\caption{(Color online)  Magneto-optical effects of MnF$_2$ with $\mathbf{n}\parallel$ [001].  Real (a) and imaginary (b) parts of the diagonal optical conductivity, and real (c) and imaginary (d) parts of the off-diagonal optical conductivity, respectively.  (e) and (f) show the Kerr and Faraday rotation angles, respectively.  The blue and red solid lines indicate strain applied along the $ac$ ($\varepsilon_{ac}$) and $ab$ ($\varepsilon_{ab}$) directions, respectively, and the color gradient from dark to light denotes increasing strain amplitude.  For comparison, the magneto-optical responses of the pristine MnF$_2$ ($\varepsilon_{no}$) are also shown as black lines.}
		\label{fig:MOKE}
	\end{figure}

	Finally, we turn to the strain-induced magneto-optical responses of semiconducting MnF$_2$, based on the Eqs.~\ref{eq:MOK}--\ref{eq:OPC}.  The calculated optical conductivities ($\sigma_0$, $\sigma_{xy}$, and $\sigma_{zx}$) for the $\varepsilon_{no}$, $\varepsilon_{ac}$, and $\varepsilon_{ab}$ cases are displayed in Figs.~\ref{fig:MOKE}(a-d).  Here, $\sigma_0$ is the averaged diagonal optical conductivity in the relevant polarization plane.  Specifically, for $\varepsilon_{ac}$ and $\varepsilon_{ab}$, we define $\sigma_0 = (\sigma_{zz} + \sigma_{xx})/2$ and  $\sigma_0 = (\sigma_{xx} + \sigma_{yy})/2$, respectively.  The absorptive parts $\sigma_{0}'$ and $\sigma_{xy}''$ (or $\sigma_{zx}''$) measure the average and difference in the absorption of left- and right-circularly polarized light, respectively.  Compared with the pristine case $\varepsilon_{no}$, $\sigma_{0}'$ changes only slightly for both $\varepsilon_{ac}$ and $\varepsilon_{ab}$, with no significant modification of the overall spectral line shape.  This weak strain dependence can be understood from the fact that the relatively small shear strain considered here only mildly perturbs the band structure and therefore have a limited influence on the interband transition probability and the joint density of states, to which $\sigma_{0}'$ is directly related.  To further identify the microscopic origins of the main absorption peaks in Fig.~\ref{fig:MOKE}(a), we determine the symmetry-allowed interband transitions based on the dipole selection rules at the X point and along the relevant high-symmetry path. For the no strain case, the interband transitions $\mathrm{X}_3 \rightarrow \mathrm{X}_4$, $\mathrm{X}_4 \rightarrow \mathrm{X}_3$, $\mathrm{X}_3 \rightarrow \mathrm{X}_3$, and $\mathrm{X}_4 \rightarrow \mathrm{X}_4$ are symmetry-allowed at the X point, and the $\mathrm{D}_5 \rightarrow \mathrm{D}_5$ transition is allowed along the X-M path. Accordingly, the $\mathrm{A}_1$ peak at 6.27 eV originates from the interband transition $\mathrm{X}_3 \rightarrow \mathrm{X}_4$, whereas the $\mathrm{A}_2$ and $\mathrm{A}_3$ peaks at 7.01 eV and 7.48 eV, respectively, originate from the interband transitions $\mathrm{D}_5 \rightarrow \mathrm{D}_5$, as shown in Fig.~\cite{SuppMater}.  By contrast, the $\sigma_{xy}$ and $\sigma_{zx}$ are substantially modified by the strain engineering, as shown in Figs.~\ref{fig:MOKE}(c) and \ref{fig:MOKE}(d).   The spectra of $\sigma_{xy}''$ and $\sigma_{zx}''$ exhibit pronounced oscillations, and their positive and negative values indicate that the corresponding interband transitions are predominantly driven by left- and right-circularly polarized light, respectively.  The corresponding dispersive components, $\sigma_{0}''$, $\sigma_{xy}'$ (or $\sigma_{zx}'$), can then be obtained from the absorptive parts through the Kramers-Kronig transformation~\cite{Bennett1965_PR}.

	After obtaining the optical conductivity tensors, we can calculate the MOKE and MOFE.  The Kerr and Faraday rotation angles ($\theta_K$ and $\theta_F$) as a function of photon energy are illustrated in Figs.~\ref{fig:MOKE}(e) and \ref{fig:MOKE}(f), respectively.  As expected for semiconducting MnF$_2$, both the $\theta_K$ and $\theta_F$ vanish within the band gap.  Once the photon energy exceeds the band gap, both $\theta_K$ and $\theta_F$ exhibit pronounced oscillations, and their magnitudes generally increase with increasing strain amplitude.  For $\varepsilon_{ab}$, the maximum values of $\theta_K$ and $\theta_F$ reach approximately 0.1 deg and 3 deg/$\mu$m.  Notably, the strain direction can even reverse the signs of $\theta_K$ and $\theta_F$.  This behavior originates from the strain-selective activation of different off-diagonal optical conductivity components, namely $\sigma_{xy}$ for $\varepsilon_{ab}$ and $\sigma_{zx}$ for $\varepsilon_{ac}$.

	\section{Summary}
	In summary, based on the minimal TB model, we demonstrate that directional shear strain provides an effective and experimentally accessible route to control magnetic phase transitions in altermagnets.  Depending on the strain direction, the system can either remain altermagnetic or evolve into a ferrimagnetic descendant phase, namely a partially compensated ferrimagnet in metals and a fully compensated ferrimagnet in semiconductors.  These symmetry-governed strain effects are further verified by spin-space group analysis and first-principles calculations for the realistic altermagnets RuO$_2$ and MnF$_2$.  The nonrelativistic spin-resolved band structures, together with the TDOS and IDOS, show that the system remains altermagnetic with shear strain applied along the $ac$ direction, whereas with shear strain applied along the $ab$ direction, RuO$_2$ evolves into a partially compensated ferrimagnetic metal because of its metallic character, while MnF$_2$ becomes a fully compensated ferrimagnetic semiconductor owing to its semiconducting gap.  Through magnetic-group analysis, we further reveal that strain lowers the crystal symmetry and thereby enables anomalous transport and magneto-optical responses, including the AHE, ANE, ATHC, MOKE, and MOFE, which are forbidden in the pristine systems.  First-principles calculations show that these responses exhibit a pronounced strain dependence, and the AHC, ANC, ATHC, as well as the magneto-optical rotation angles, all increase with increasing strain amplitude.  Our work establishes a general strategy for driving altermagnets into fully or partially compensated ferrimagnetic phases, deepens the understanding of unconventional antiferromagnets, and broadens their potential for antiferromagnetic spintronics and magneto-optical device applications.
	
	\begin{acknowledgments}
		The authors thank Jingyi Duan, and Chaoxi Cui for their helpful discussion.  This work is supported by the National Natural Science Foundation of China (Grants No. 12404052, No. 12304066,  No. 12474012, No. 12404256), the Basic Research Program of Jiangsu (Grant No. BK20241049 and No. BK20230684), the Natural Science Fund for Colleges and Universities in Jiangsu Province (Grant No. 24KJB140011 and No. 23KJB140008).
		
	\end{acknowledgments}
	
	\bibliography{references}

\end{document}